\documentclass[a4paper,11pt]{article}
\usepackage{eepic,epsfig,pstricks}
\usepackage{amsmath}
\usepackage{amsfonts}      
\newcounter{num}

\def\Ucgm{U_{\chi,g^{-1}}}
\def\Uc#1{U_{\chi,{#1}}}

\begin{document}
%
\newcommand{\CAL}{Conditions aux limites }
\newcommand{\Cal}{conditions aux limites }
\newcommand{\cad}{c'est-\`a-dire }
\newcommand{\Pc}{\ensuremath{\widehat{P}}}
\newcommand{\Gc}{\ensuremath{\widehat{G}}}
\newcommand{\Ucg}{\ensuremath{U_{\chi,g}}}
\newcommand{\Ucbg}{\ensuremath{U_{\overline{\chi},g}}}
\newcommand{\Hc}{\ensuremath{\mathcal{H}}}
\renewcommand{\phi}{\ensuremath{\varphi}}
\newcommand{\phib}{\ensuremath{\overline{\varphi}}}
\newcommand{\tem}{temp\'erature }
\newcommand{\tems}{temp\'eratures }
\newcommand{\nab}{\vec{\nabla}}
\newcommand{\Ui}{\vec{U_i}}
\newcommand{\nabUi}{\vec{\nabla}.\vec{U_i}}
\newcommand{\Us}{\vec{U_s}}
\newcommand{\V}{\vec{V}}
\newcommand{\Vt}{\vec{\tilde{V}}}
\newcommand{\taua}{\vec{\tau_a}}
\newcommand{\taus}{\vec{\tau_s}}
\newcommand{\FI}{\vec{F_I}}
\newcommand{\Fc}{\vec{F_c}}
\newcommand{\F}{\vec{F}}
\newcommand{\NF}{\|\vec{F}\|}
\newcommand{\rw}{\rho_w}
\newcommand{\ra}{\rho_a}
\newcommand{\CD}{C_D}
\newcommand{\pmax}{p_{max}}
\newcommand{\pmaxij}{p_{max,ij}}
\newcommand{\gradp}{\nab p}
\newcommand{\gradpt}{\nab \tilde{p}}
\newcommand{\pt}{\tilde{p}}
\newcommand{\Dt}{\tilde{D}}
\newcommand{\R}{\overline{\overline{R}}}
\newcommand{\M}{\overline{\overline{M}}}
\newcommand{\Rmalpha}{\overline{\overline{R}}_{(-\alpha)}}
\newcommand{\Rmtalpha}{\overline{\overline{R}}_{(-2\alpha)}}
\newcommand{\sinmalpha}{sin(-\alpha)}
\newcommand{\dlt}{\delta t}
\newcommand{\Xrz}{X_{r}}
\newcommand{\Xr}[1]{X_{r(#1)}}
\newcommand{\Xrpz}{X_{rp}}
\newcommand{\Xrp}[1]{X_{rp(#1)}}
\newcommand{\Xrvz}{X_{rv}}
\newcommand{\Xrv}[1]{X_{rv(#1)}}
\newcommand{\Xrhz}{\hat{X}_{r}}
\newcommand{\Xrh}[1]{\hat{X}_{r(#1)}}
\newcommand{\MM}{{\cal M}}
\newcommand{\DD}{{\cal D}}
\newcommand{\IP}{{\cal I}_+}
\newcommand{\IM}{{\cal I}_-}
\newcommand{\dsdlt}{\frac{2}{\dlt}}
\newcommand{\vphi}{\vec{\varphi}}
\newcommand{\veta}{\vec{\eta}}
\newcommand{\veps}{\vec{\varepsilon}}
\newcommand{\deleta}{\vec{\delta \eta}}
\newcommand{\deleps}{\vec{\delta \varepsilon}}
\newcommand{\delphi}{\vec{\delta \varphi}}
\newcommand{\vG}{\vec{G}}
\newcommand{\vf}{\vec{f}}
\newcommand{\n}{\vec{n}}
\newcommand{\rl}{\rho_{l}}
\newcommand{\rv}{\rho_{v}}
\newcommand{\vv}{\vec{V}_{v}}
\newcommand{\vl}{\vec{V}_{l}}
\newcommand{\g}{\vec{g}}
\newcommand{\ul}{u_{l}}
\newcommand{\uv}{u_{v}}
\newcommand{\Cl}{C_{l}}
\newcommand{\dsdt}{\frac {\partial }{\partial t}}


\begin{center}

\vspace*{2cm}

\LARGE{Particle description of zero energy vacuum.\\
        I. Virtual particles}

\vspace{5mm}

\normalsize{Jean-Yves Grandpeix\,\footnotemark[1], Fran\c{c}ois Lur\c{c}at\,\footnotemark[2]}

\vspace{3cm}

{\bf Abstract}
\end{center}

\medskip

First the "frame problem" is sketched: the motion of an isolated particle obeys a simple law in 
galilean frames, but how does the galilean character of the frame manifest itself at the place 
of the particle? A description of vacuum as a system of virtual particles will help to answer 
this question. For future application to such a description, the notion of global particle is 
defined and studied. To this end, a systematic use of the Fourier transformation on the 
Poincar\'e  group is needed. The state of a system of n free particles is represented by a 
statistical operator W, which defines an operator-valued measure on $\Pc^n$ ($\Pc$ is the 
dual 
of the Poincar\'e  group). The inverse Fourier-Stieltjes transform of that measure is called the 
characteristic function of the system; it is a function on $P^n$. The main notion is that of 
global characteristic function: it is the restriction of the characteristic function to the 
diagonal subgroup of $P^n$; it represents the state of the system, considered as a single 
particle. The main properties of characteristic functions, and particularly of global 
characteristic functions, are studied. A mathematical Appendix defines two functional spaces 
involved.

\footnotetext[1]{Laboratoire de M\'et\'eorologie Dynamique,
  C.N.R.S.-Universit\'e Paris 6,
 F-75252 Paris Cedex 05, France.
 E-mail: jyg @lmd.jussieu.fr}
\footnotetext[2]{Laboratoire de Physique Th\'eorique (Unit\'e Mixte de Recherche (CNRS) UMR 8627).
 Universit\'e de Paris XI, B\^atiment 210,
 F-91405 Orsay Cedex, France.
 E-mail: francois.lurcat@wanadoo.fr}


\newpage

\section{INTRODUCTION}

In this paper and in the companion article(1), we present a description of the vacuum as a 
system of virtual particles. During the last half-century or so, virtual particles have been 
stowaways in theoretical physics; we intend to show that they deserve to have a better status. 
It will turn out that their state can be described quantum-mechanically, at the price of a 
natural extension of the usual density matrix formalism. The extended formalism will then be 
used to give an explicit description of the vacuum in terms of virtual particles. It will be 
expedient to use a mathematical tool which may be not so familiar to physicists, namely 
harmonic analysis on the Poincar\'e  group; but as far as physics is concerned, we shall be 
elementary.

\section{THE FRAME PROBLEM}

In classical mechanics (non-relativistic or relativistic) the motion of an isolated particle 
has simple kinematical properties summarized by the law of inertia, which holds true only in 
galilean frames. In quantum mechanics it takes the form of the law of conservation of momentum 
for an isolated particle, again restricted to galilean frames. But neither classical mechanics, 
nor non-relativistic quantum mechanics give any sensible solution to the {\it frame problem: 
how does the galilean character of the frame manifest itself physically at the place of the 
particle?} In textbooks and treatises it is systematically avoided, and students are taught to 
refrain from asking about it.

 On the other hand, relativistic quantum mechanics gives an essentially 
different description of the inertial motion, which fundamentally solves the frame problem. 
Already in the Dirac equation there are clues to that, namely the {\it Zitterbewegung} and the 
fact 
that the only eigenvalues of the velocity operator are $\pm c$. As stressed with particular 
strength by Weinberg(2), however, any relativistic quantum theory is equivalent to a quantum 
field theory\,\footnotemark[3].

\footnotetext[3]{More precisely, Weinberg states that if one lets aside string and similar theories, 
quantum field theory is the only way to reconcile the principles of quantum mechanics 
(including the cluster decomposition property) with those of relativity.}

Now relativistic quantum field theory does not recognize permanent particles: it works from 
start to finish with the creation and annihilation of particles. This should not be forgotten 
even when dealing  with inertial motion. Frenkel, who made this remark long ago(3), described the 
inertial motion of an electron as a sequence of processes of the following type: annihilation 
of the electron with the positron of a virtual pair, and transition of the virtual electron to 
a real state.

A similar description was given by Thirring in his textbook(4). It may be worth 
noting that Rosenfeld severely criticized this approach(5): he insisted that virtual pairs 
should be considered as nothing more than "easily remembered semi-quantitative features of the 
formalism". While his remark rightly warns us against a too literal interpretation of the 
Frenkel-Thirring process, we shall take the essential physical idea of this process as an 
incentive (among others) to build a rigorous description of vacuum as a system of virtual 
particles.

\subsection{Vacuum Virtual Particles as a Possible Solution of the Frame Problem}

Our central assumption is that vacuum is a system of virtual particles. This will bring us to 
consider the inertial motion as a guided motion: a process of permanent interaction with the 
virtual particles of vacuum. 
The relativistic invariance of vacuum implies, as will be seen in (II), that the virtual 
particles are homogeneously distributed. Accordingly, the momentum of a free particle (i.e. a 
particle interacting only with the vacuum distribution) is conserved as it proceeds. Such is 
the physical idea of the solution of the frame problem.

 Now what about non-galilean frames? In such a frame, the motion of a particle is again a 
process of interaction with the virtual particles of vacuum. But now momentum is no longer 
conserved, hence the distribution of the vacuum particles is no longer homogeneous. We shall 
say that vacuum is deformed, and that the accelerated motion of a particle is due to its 
interaction with the deformed vacuum.

\subsection{The Semantic Incompleteness of Non-vacuum Theories}

Maybe this kind of ideas could help to deal with physical problems at the boundary between 
gravitation and quantum mechanics. Ignoring this possibility for the time being, let us merely 
remark that our reformulation cures theoretical physics from an illness common to classical 
mechanics and non-relativistic quantum mechanics: in so far as they cannot solve the frame 
problem, their descriptions of motion and of gravitational field bear a strongly formal 
character. Physics indeed tries to explain the phenomena which it considers by interactions 
between material objects. But in theories which cannot take into account the existence of 
vacuum, there is no physical interaction between the reference frame and the moving particle. 
Hence, even though the difference between galilean and non-galilean frames is exactly described 
(above all in general relativity), it lacks {\it Anschaulichkeit}. (Or to tell it in Feynman's 
terms(6)(7), it hardly allows visualization). Of course this does not lead to any logical 
incompleteness; but these theories show a deficit of intelligibility. With growing habit 
physicists no longer notice the deficit, which is sometimes better felt by students; still one 
has to admit that non-vacuum theories suffer from a semantic incompleteness.

\subsection{Resonances as Virtual Particles}
 \label{sec.2.3}

Another reason to take more seriously the concept of virtual particles can be found in the 
phenomenology of strong interactions. It was found indeed very early, even before the advent of 
quarks, that resonances are not only peculiarities of interactions between stable particles, 
but also full-fledged particles. But what about their status: are they real or virtual 
particles? A virtual particle is one which has too short a lifetime to be detectable 
"directly", i.e., by its interaction with a macroscopic apparatus. It can be detected only 
through the mediation of its interaction with some real particles. This definition is usually 
accepted, most often implicitly. It implies immediately that resonances are virtual particles. 
As an almost trivial corollary, there should be no shame in speaking about virtual particles, 
not only as ways of describing a formalism, but also as physical objects. Hence the 
quantum-mechanical description of the state of a resonance(8) should be considered as a matter 
of course; that this is not the case presently is one more clue to the unreasonable status of 
virtual particles in current physics.

\section{THE FOURIER TRANSFORMATION}

In the following we shall proceed up to a characterization of the state of virtual particles. A 
virtual particle is an object present in the intermediate stage of a process. In the case of a 
formation reaction, the initial (real) particles transform into the virtual particle of 
interest; hence we shall define the state of the virtual particle as the global state of the 
initial particles (or equivalently, of the final particles). By global state of a system of 
(real) particles, we understand(9) the state of that system considered as a single particle; of 
course this is a natural generalization of the notion of centre-of-mass motion. We shall also 
say that the global state is the state of the global particle of the system.

\subsection{Why the Fourier Transformation?}

In order to define the state of the global particle of a given system of particles, we shall 
have to introduce Fourier transformation on the invariance group.

 This transformation will turn out to be a natural tool of quantum relativistic kinematics. By 
the usual Fourier transformation, a function is represented as a combination of exponentials. 
Why exponentials, and not another family of functions? Because the function $e^{i\omega t}$ 
defines a unitary representation of the translation group. In the frequency analysis of a 
signal, there is indeed an implicit reference to the signal passing through a linear filter, 
invariant by the translation group(10).

In a physical theory invariant by a group $G$ larger than the translation group, one should 
therefore expect to meet the Fourier transformation on $G$. The Fourier transformation on a 
group has been long known to mathematicians(11)(12). Up to now, however, physicists hardly used 
the resources of this mathematical theory. In the following we shall show that it is a natural 
and necessary tool of quantum relativistic kinematics.

First some basic notions of Fourier transformation will be recalled. Then the characteristic 
function(8)(9) of a one-particle state will be defined: it is a function on the Poincar\'e  group 
$P$, which turns out to be the inverse Fourier-Stieltjes transform of the operator-valued 
measure on the dual $\Pc$, defined by the statistical operator (usually called "density 
matrix") of the state.

The characteristic function of an n particle state will be defined in an analogous manner: it 
is a function on $P^n$. The restriction of this function to the diagonal subgroup of $P^n$ is 
the global characteristic function (9) of the state. Its physical meaning is simple and 
important: it characterizes the global state of the system, i.e. the state of the $n$ 
particles, considered as a single particle. This notion is an elementary quantum mechanical 
one; particle physics uses it implicitly, especially in the study of resonances in production 
reactions, but it does not seem to have ever been explicitly defined. 

An Appendix shows that the two types of characteristic functions which have been introduced 
belong respectively to two mathematical objects - the Fourier algebra and the Fourier-Stieltjes 
algebra - which play an important role in harmonic analysis.

\subsection{The Fourier Transform of a Function}

The mathematical properties stated in this article are meant to be applied to the case where 
$G$ is $P$, the inhomogeneous $SL(2,C)$, or universal covering group of the Poincar\'e  group, or 
else a direct power of this group.  They will be stated mostly under a rather general form; the 
groups $P$ and $P^n$, as well as their extensions to discrete and internal symmetries, belong 
to the class of groups which will be considered. (For brevity's sake, $P$ will sometimes be 
called, somewhat improperly, the "Poincar\'e  group"). 

Let $G$ be a separable unimodular type I Lie group(13). The invariant measure on $G$ (Haar 
measure) is defined up to a factor; this factor will be choosed once for all, and the so fixed 
Haar measure will be denoted by $dg$. 

The set of the equivalence classes of irreducible continuous unitary representations of $G$ is 
called the dual of $G$, denoted by $\Gc$. (In the following, we shall say "representation" 
instead of "unitary continuous representation"). In each equivalence class we shall choose a 
certain representation, hence $\Gc$ will be considered as a set of representations. A 
representation will be labeled, in the case $m^2>0$, by $\chi = (m, s, \epsilon)$, $\epsilon$ 
being the sign of energy; this $\chi$ will be called the signature of the representation. In 
the representation $\chi \in  \Gc$, the element $g \in  G$ is represented by the unitary 
operator $\Ucg$ ; the space of this representation is denoted by $\mathcal{H}_{\chi}$.

The Plancherel measure on $\Gc$ will be denoted by $d\chi$; its normalization is fixed as soon 
as the normalization of the Haar measure is chosen. 

The definition of an operator field on $\Gc$ amounts to giving, for almost all $\chi \in  \Gc$, 
an operator $A_{\chi}$ on the space $\mathcal{H}_{\chi}$. Let $\varphi$ be a complex valued 
function on $G$. If $\varphi$ is integrable, the Fourier 
transform of $\varphi$  is the operator field on $\Gc$ defined by
\begin{equation}
 \label{eq.3.1}
   \mathcal{T}(\varphi)_{\chi} = \int_G \varphi_g \Ucg dg
\end{equation}
(For reasons to appear later, the names of variables will be written as indices). 

If the function $ \chi \rightarrow Tr[\mathcal{T}(\varphi)_{\chi} \Ucg]$ is integrable on $\Gc$ 
for any $g \in G$, equation (\ref{eq.3.1}) can be inverted:
\begin{equation}
 \label{eq.3.2}
 \varphi_g = \int_{\Gc} Tr[\mathcal{T}(\varphi)_{\chi} \Uc{g^{-1}}] d\chi
\end{equation}

\subsection{The Inverse Fourier-Stieltjes Transform of an Operator-valued Measure}

Mathematicians usually start with a function on a group and define its Fourier transform, which 
is an operator field on the dual; the inverse Fourier transformation allows one to come back to 
the function. But for the problems of interest here the starting point is an object on the 
dual. The statistical operator which describes the state of a particle is indeed related to an 
irreducible representation of the invariance group, and therefore to a point of the dual. We 
shall also introduce states which have a mass spectrum (the global states); here we shall have 
an operator field on the dual. The notion which includes these two different situations is an 
operator-valued measure on the dual, which has an inverse Fourier-Stieltjes transform (9)(14). 
Let us now proceed to define these notions. 

Let $\mu$ be a positive scalar measure on $\widehat{G}$, and let $F$ be an operator field 
defined on 
the support of $\mu$: to any $\chi \in supp(\mu)$ there corresponds an operator $F_{\chi}$ on 
$\mathcal{H}_{\chi}$.  The couple $\mathcal{M} = (\mu,F) $
defines an operator-valued measure on $\widehat{G}$. If the operator $F_{\chi}$ is positive and 
trace class $\mu$-almost everywhere, and if the function 
$\chi \rightarrow Tr(F_{\chi})$ is $\mu$-integrable, one says that the operator-valued measure 
$\mathcal{M}$ is {\it trace class}. 

If $\mathcal{M} = (\mu,F) $ is a trace class operator-valued measure on $G$ , its inverse 
Fourier-Stieltjes transform $\varphi$ is defined by a Stieltjes integral:
\begin{equation}
 \label{eq.3.3}
 \varphi_g = \int_{\Gc} Tr[F_{\chi} \Ucgm] d\mu(\chi)
\end{equation}

If the measure $\mu$ is absolutely continuous with respect to the Plancherel measure, one can 
assume without restriction that it is equal to the Plancherel measure; the operator field $F$
then appears as the {\it spectral density} of the operator-valued measure $\mathcal{M}$ with 
respect to the 
Plancherel measure (otherwise stated, as its Radon-Nikodym derivative). If $\mu$ is equal to 
the Plancherel measure, definition (\ref{eq.3.3}) becomes
\begin{equation}
 \label{eq.3.4}
  \varphi_g = \int_{\Gc} Tr[F_{\chi} \Ucgm] d\chi
\end{equation}
and we get back formula (\ref{eq.3.2}).
 
We shall need also another particular case: the case when $\mu$ is a Dirac measure at the 
point 
$\chi \in  \Gc$ , whereas the field $F$ reduces to a single positive trace class operator $W$ 
on $\mathcal{H}_{\chi}$. The 
operator-valued measure thus defined will be called the Dirac measure $W$ at the point $\chi$. 
Its inverse Fourier-Stieltjes transform is given by
\begin{equation}
 \label{eq.3.5}
 \varphi_g = Tr[W \Ucgm]
\end{equation}

\subsection{Examples}

{\bf  $SU(2)$}

For a compact group such as $SU(2)$, the dual is discrete; in Eq. (\ref{eq.3.2}), the 
integration reduces to 
a summation. The natural normalisation of $dg$ is defined by putting the total measure of the 
group equal to $1$; as a result, the Plancherel measure of a representation is equal to its 
dimension. 

For $SU(2)$, the Haar measure is defined by
$$
\int_G f_g dg = (8 \pi^2)^{-1} \int_0^{2 \pi} d\alpha \int_0^{\pi} sin \beta d\beta \int_0^{2 \pi} d\gamma f(\alpha,\beta,\gamma)
$$
where $\alpha$, $\beta$, $\gamma$ are the Euler angles.

The dual can be labelled by the set of the possible values of the angular momentum:
$$
  \Gc  = \{0,1/2,1,3/2,...\}
$$
The Plancherel measure consists in masses $2j+1$ at the points $j$:
\begin{equation}
 \label{eq.3.6}
  p_j = 2j+1
\end{equation}

For any $j \in \Gc$ , the space $\Hc_j$ has dimension $2j+1$ and the operator $U_{j,g}$ is 
defined by the matrix 
$$
    <m|U_{j,g}|m'> = D^j_{mm'}(g)
$$
where the $D^j_{mm'}$ are the well known matrix elements of the $j$ representation of $SU(2)$. 
The 
Fourier transformation sends the function $f$ on $G$ onto the operator field defined thus: to 
any $j$ corresponds the operator
\begin{equation}
 \label{eq.3.7}
  \mathcal{T}(f)_j = \int_G f_g U_{j,g} dg \;.
\end{equation}

The Fourier inversion formula expresses the function $f$ in terms of the operator field 
$\mathcal{T}(f)$:
\begin{equation}
 \label{eq.3.8}
   f_g = \sum_j p_j Tr[\mathcal{T}(f)_j U_{j,g^{-1}} ]
\end{equation}
It can be obtained directly by using the orthogonality relations of the matrix elements of the 
representations of $SU(2)$. The form of equation (\ref{eq.3.8}) makes obvious the fact that it 
is a particular case of equation (\ref{eq.3.2}). 

But we shall need, rather than the inverse Fourier transform (\ref{eq.3.8}), the inverse 
Fourier-Stieltjes transform of an operator-valued measure, especially that of the Dirac measure 
$W$ at the point $j$. This transform is a particular case of Eq. (\ref{eq.3.5}): 
\begin{equation}
 \label{eq.3.9}
  f_g = Tr[W U_{j,g^{-1}} ]
\end{equation}

{\bf Poincar\'e group $P$}

The above defined group $P$ is locally compact, but not compact: its total Haar measure is 
infinite. Hence the dual is not discrete. In physics one uses most often the part of the dual 
which corresponds to the irreducible representations with positive squared mass and energy. Let 
us remark at once that the representations which differ from the ones just mentioned only by 
having a negative energy are as simple, they can be obtained for instance by taking the complex 
conjugate of the operator for positive energy. These two classes of representations are 
relevant for the study of vacuum (see the companion paper (II)); hence their products (which 
contain some representations with $m^2<0$) are relevant also. The following considerations hold 
for all the mentioned classes of representations. However, we shall always take as examples the 
well known representations with $m^2>0$ and positive energy. 

In order to define the Haar measure and to fix its normalization, let us write with Nghi\^em(15)(16)
an element of $SL(2,C)$ under the canonical form
$$
 A = U H_k
$$
 In this equation $U$ denotes an 
element of $SU(2)$; $H_k$ is the unique hermitian positive matrix which sends the vector $k$ 
onto the 
vector $(1,0,0,0)$; finally, $k$ is the image of the vector $(1,0,0,0)$ by $A^{-1}$. We can now 
define the Haar measure on $SL(2,C)$: it is given by
$$
  d^6 A = d^3 U d^3 \mathbf{k} / k_0  \;,
$$
where $d^3 U$ is the Haar measure on $SU(2)$. Hence the Haar measure on $P$ is given by
\begin{equation}
 \label{eq.3.10}
  dg = d^4 a d^6 A
\end{equation}

This normalization of the Haar measure fixes that of the Plancherel measure. In particular, 
the latter reads on the representations with positive $m^2$ thus ($s$ denotes the spin value):
\begin{equation}
 \label{eq.3.11}
  d\chi = [2(2\pi)^4]^{-1} (2s+1) m^2 dm^2
\end{equation}

The Fourier transform of a function on $P$ is the operator field on $\widehat{P}$ defined by 
equation (\ref{eq.3.1}). 
The Fourier inversion formula is given by equation (\ref{eq.3.2}). The extension to the case 
$G=P^n$ is straightforward. 

Practically, to compute the operator $\mathcal{T}(\varphi)_{\chi}$ and to evaluate the trace in 
equation (\ref{eq.3.2}) one uses 
an improper basis in the space $\Hc_{\chi}$. See Ref. 9, and for more details Nghi\^em's 
papers(15)(16). 

Finally, the inverse Fourier-Stieltjes transform of the operator-valued Dirac measure $W$ at 
the point $\chi$ is given by equation (\ref{eq.3.5}).

\section{CHARACTERISTIC FUNCTIONS}

\subsection{One Particle}

Let us first consider a spin $j$, i.e. a particle of spin $j$, of which we consider only the 
spin 
degrees of freedom. Its state is defined by a statistical operator $W$ on the space $\Hc_j$ 
(whose 
dimension is $2j+1$) of the representation $j$ of $SU(2)$. To this state one may also associate 
the Dirac operator-valued measure $W$ at the point $j$. 

We define the {\it characteristic function} of the state as the inverse Fourier-Stieltjes 
transform 
of the latter operator-valued measure. This transform is the function $\varphi$ given by 
equation 
(\ref{eq.3.9}). As $W$ is trace class, the function is defined everywhere on the group; the 
trace 
of $W$ is equal to the value of the characteristic function at the neutral element of the 
group:
\begin{equation}
 \label{eq.4.1}
  Tr W = \varphi_e
\end{equation}
The positivity of the operator $W$ is equivalent to the fact that $\varphi$ is a 
positive-definite function(13)(17). 
It should be noticed that the characteristic function is not the inverse Fourier transform of 
the operator field defined by putting $W$ at the point $j$, zero elsewhere; the latter is given 
by 
Eq. (\ref{eq.3.8}), not Eq. (\ref{eq.3.9}); equation (\ref{eq.3.8}) defines the same function, 
but multiplied by $(2j+1)$. The choice made extends to a non-compact group. 

Let us now consider a particle transforming according to the representation $\chi$ of the 
group $G=P$. Its state is represented by a statistical operator $W$ on the space $\Hc_{\chi}$. 
The associated 
operator-valued measure is the Dirac measure $W$ at the point $\chi$. Again, we define the 
characteristic function of the state as the inverse Fourier-Stieltjes transform of the latter 
operator-valued measure. It is given by equation (\ref{eq.3.5}). As $W$ is trace class, this 
function is defined everywhere on $G$. 

As for a spin, the positivity of $W$ is equivalent to the fact that $\varphi$ is a 
positive-definite 
function. The trace of $W$ is again given by equation (\ref{eq.4.1}), where $e$ denotes the 
neutral element of $G$.

\subsection{Several Particles}

Let us now consider the case of $n$ spins $j_1,...,j_n$. The state of the system is represented 
by 
an operator $W$ on the space
$$
   \Hc_{j_1 ... j_n} = \Hc_{j_1}\otimes ... \otimes \Hc_{j_n} \; .
$$
 This case is analogous to that of a single spin: one should only 
replace the group $SU(2)$ by $SU(2)^n$, a group whose elements read $(g_1... g_n)$. The 
elements of the dual of $SU(2)^n$ read $(j_1,..., j_n)$; the operator $U_{j_1 ... j_n,g_1 ... g_n}$
 is the tensor product
$$
   U_{j_1 ... j_n,g_1 ... g_n} = U_{j_1,g_1} \otimes ... \otimes U_{j_n,g_n}  \; .
$$
The characteristic function of the state $W$ reads
\begin{equation}
 \label{eq.4.2}
  \varphi_{g_1 ... g_n} = Tr \{ W  U_{j_1 ... j_n,g_1^{-1} ... g_n^{-1}} \}
\end{equation}

As in the case $n=1$, this equation defines the inverse Fourier-Stieltjes transform of an 
operator-valued measure, namely the Dirac measure $W$ at the point $(j_1,...,j_n)$ of the dual. 

Let us consider now $n$ particles, corresponding to the irreducible representations 
$\chi_1, ... ,\chi_n$ 
of the group $P$. The state of the system is represented by an operator $W$ on the space
$$
   \Hc_{\chi_1 ... \chi_n} = \Hc_{\chi_1}\otimes ... \otimes \Hc_{\chi_n} \; .
$$ 

This case is analogous to that of a single particle: one should only replace the group $P$ by 
$P^n$, 
a group whose elements read $(g_1...g_n$). The elements of the dual of $P^n$ read 
$(\chi_1, ...,\chi_n)$; the operator $U_{\chi_1 ... \chi_n,g_1 ... g_n}$ is the tensor 
product
\begin{equation}
 \label{eq.4.3}
   U_{\chi_1 ... \chi_n,g_1 ... g_n} = U_{\chi_1,g_1} \otimes ... \otimes U_{\chi_n,g_n}  \; .
\end{equation}

The characteristic function of the state $W$ reads
\begin{equation}
 \label{eq.4.4}
    \varphi_{g_1 ... g_n} = Tr \{ W  U_{\chi_1 ... \chi_n,g_1^{-1} ... g_n^{-1}} \}
\end{equation}

As in the case $n=1$, this equation defines the inverse Fourier-Stieltjes transform of an 
operator-valued measure, namely the Dirac measure $W$ at the point $(\chi_1, ... ,\chi_n)$ of 
the dual.

\section{PROPERTIES OF THE CHARACTERISTIC FUNCTIONS}

The properties and the physical meaning of characteristic functions have been studied in detail 
in Ref. 9. Let us recall here some essential results.

\subsection{Meaning of the Characteristic Function}

 As shown by equation (\ref{eq.3.5}), the value of the characteristic function $\varphi$ of a 
state at 
the element g of the group is the expectation value in this state of the operator 
$\Uc{g^{-1}}$. This 
definition is analogous to that of the characteristic function of a random variable $X$: its 
value at $t$ is the expectation value of $e^{itX}$. Furthermore in the case $G=P$, the 
restriction of $\varphi$ 
to the translation subgroup, $a \rightarrow \varphi(a,e)$, turns out to be the characteristic 
function (in the 
usual sense of probability theory) of the energy-momentum vector. If indeed we write an element 
of $P$ under the form $(a,A)$ with $a \in T$ (the translation subgroup) and $A \in  SL(2,C)$, 
the operator 
$U(a,e)$ can be written $exp(ia_{\mu}P^{\mu})$. (Both the neutral elements of $P$ and of 
$SL(2,C)$ will be 
denoted by $e$). Hence the expectation values of the components of the energy-momentum vector 
are given by  
\begin{equation}
 \label{eq.5.1}
    < P^{\mu} > = [i/\phi_e] \partial \phi_{a,e}/\partial a_{\mu} |_{a=0}  \quad .
\end{equation}

(We did not assume that the statistical operator is normalized; we have used equation 
(\ref{eq.4.1}). 

More generally, the expectation values of the dynamical variables of the system can be obtained 
by applying to the characteristic function suitable differential operators, which correspond to 
elements of the universal enveloping algebra of the Lie algebra of the Poincar\'e  group. 

Let us define the partial Fourier transform(16) (on the translation subgroup) of the 
characteristic function $\phi$:
$$
  \phib_{k,A} = \int e^{ika} \phi_{a,A} d^4a  \quad .
$$
As $\varphi(a,e)$ is the characteristic function (in the probabilistic sense) of the 
energy-momentum 
vector, the restriction of $\overline \varphi$ defined by putting $A=e$ is the (unnormalized) 
probability density of the energy-momentum vector. 
Let us now consider the restriction of $\phib$ defined by giving to $k$ a fixed value and 
restraining $A$ 
to the little group of $k$, i.e. to the subgroup of the elements of $SL(2,C)$ which leave $k$ 
fixed: 
$Ak=k$. It can be shown(9), thanks to Nghi\^em's results(16), that this restriction gives the 
spin 
characteristic function (cf. Section \ref{sec.2.3}) of the particle. Explicitly, the spin 
characteristic function is given by
\begin{equation}
 \label{eq.5.2}
    \omega_{k,g} = \phib_{k,H_k^{-1}gH_k}
\end{equation}
where $H_k$ has been defined in Section \ref{sec.2.3}, and $g$ is an element of $SU(2)$. 

\subsection{Transformation Law}

Let us consider a one particle state defined by its statistical operator $W$ and by its 
characteristic function $\phi$; let $\chi$ be the irreducible representation associated to the 
particle. 
The element $\gamma$ of the invariance group transforms this state into another one, whose 
statistical operator is
$$
    W' = \Uc{\gamma} W \Uc{\gamma^{-1}}   \quad .
$$
Equation (\ref{eq.3.5}) then gives  the new characteristic function $\phi'$ : 
$$
   \phi^{'}_g = Tr[\Uc{\gamma} W \Uc{\gamma^{-1}} \Uc{g^{-1}} ]
             = Tr[W \Uc{\gamma^{-1}} \Uc{g^{-1}} \Uc{\gamma} ]  \quad ,
$$
and finally
$$
   \phi^{'}_g = \phi_{\gamma^{-1} g \gamma}   \quad .
$$
This transformation can be generalized to the case of an $n$ particle state. If there is no 
interaction between the particles, we might consider $P^n$ as the invariance group, because in 
this case the dynamical variables relative to each single particle are separately conserved. 
However, the interesting cases are those where there is an interaction somewhere; for instance, 
the state might be an asymptotic ingoing or outgoing state. The invariance group is then the 
diagonal subgroup $G_d$ of $G$, defined by
$$
 g_1 = ... = g_n  \quad .
$$
The transformation law reads
\begin{equation}
 \label{eq.5.3}
    \phi^{'}_{g_1 ... g_n} = \phi_{\gamma^{-1} g_1 \gamma ... \gamma^{-1} g_n \gamma }  \quad .
\end{equation}

\subsection{Inclusive Characteristic Functions}

Let us note finally that the characteristic function of a system of particles also allows one 
to express the inclusive characteristic functions. An inclusive state of a system of $n$ 
particles is obtained by neglecting some of them, for instance those numbered by $p+1,...,n$. 
One has to take the partial trace of the $n$ particle statistical operator with respect to the 
degrees of freedom corresponding to the $(n-p)$ neglected particles. This operation becomes 
very simple with characteristic functions:
\begin{equation}
 \label{eq.5.4}
    \phi^{(p)}_{g_1 ... g_p} = \phi^{(n)}_{g_1 ... g_p e ... e}  \quad .
\end{equation}
One puts equal to $e$ (the neutral element of the group) the variables corresponding to 
neglected particles.

\section{GLOBAL CHARACTERISTIC FUNCTIONS}

\subsection{The State of a Global Particle}

Let us consider a system of $n$ particles described by irreducible representations of the 
invariance group. What are the properties of this system, if we consider it as a single 
particle? (In the following, this particle will be called the {\it global particle}). This 
problem is 
often approached, although not quite explicitly. For instance, one may study the motion of the 
center of mass. One may also consider the total angular momentum of the system, or else its 
effective mass (i.e. the mass of the global system); the latter case occurs in the experimental 
study of resonances in production processes. 

To get the state of the global particle, one has to take the partial trace(18) of the 
statistical operator of the $n$ particle system over all degrees of freedom, except those which 
correspond to the global motion. This operation, however, is not always easy to carry out. Let 
us show that it is very simple if one uses characteristic functions. 

Let $\phi$ be the characteristic function of the system: it is a function on $P^n$, defined by 
\begin{equation}
 \label{eq.6.1}
    \phi_{g_1 ... g_n} = Tr \{ W [ U_{\chi_1,g_1^{-1}} \otimes ... \otimes U_{\chi_n,g_n^{-1}} ] \} 
\end{equation}

Let us now consider the restriction $\phi_{glob}$ of $\phi$ to the diagonal subgroup. Clearly, 
this 
is the characteristic function of the global particle (or, as we shall call it, the {\it global 
characteristic function}). Indeed, the restriction to the diagonal subgroup means that one 
considers only the combined motions of the system. Besides, definition (\ref{eq.6.1}) implies 
that the 
expectation values of the dynamical variables $P_{\mu}, M_{\mu \nu}$, which correspond to 
elements of the Lie 
algebra of the group, are obtained by adding the expectation values of the $n$ single particle 
variables. (See Eq. (\ref{eq.5.1}) for the variable $P_{\mu}$). 

To study the global characteristic function in more detail, let us consider the operator field 
which is its Fourier transform:
\begin{equation}
 \label{eq.6.2}
    \rho_{\chi} = \int_G \phi_{glob \; g} \Uc{g} dg  \quad .
\end{equation}
This field is the spectral density of an operator-valued measure. Let $K$ be a Borel part of 
$G$: the operator-valued measure assigns to it the operator
\begin{equation}
 \label{eq.6.3}
    W(K) = \int_K^{\oplus} \rho_{\chi} d\chi
\end{equation}
on the space
\begin{equation}
 \label{eq.6.4}
    \mathcal{H}(K) = \int_K^{\oplus} \mathcal{H}_{\chi} d\chi   \quad .
\end{equation}
The operator W(K) will be called a {\it conditional statistical operator} (it is conditioned by 
$K$). It is not normalized. 

The operator-valued measure just defined is another representation of the state of the global 
particle. The mathematical object which has an immediate physical meaning is not the 
{\it density of statistical operator} $\rho_\chi$, but rather the operator $W(K)$. The latter 
is dimensionless, as are the 
probabilities computed from it, whereas $\rho_\chi$ has the dimension $M^{-4}$ because $d\chi$ 
has the dimension 
$M^4$, see equation (\ref{eq.3.11}). On the dual another measure than the Plancherel measure 
might be used, 
for instance $dm^2$; in such a case the density of statistical operator would be modified, but 
$W(K)$ would not be. From this point of view, the expression (\ref{eq.6.3}) of the statistical 
operator 
is analogous to that of the power radiated by a source in the frequency range $[\nu_1,\nu_2]$:
$$
   W([\nu_1,\nu_2]) =\int_{\nu_1}^{\nu_2}  (dW/d\nu) d\nu  \quad .
$$
This equation expresses the fact that the radiated power is an additive function of the 
frequency interval. If instead of the frequencies one uses the wavelengths, the spectral 
density of power becomes $dW/d\lambda = (\nu^2/c) (dW/d\nu)$, but the radiated power is of 
course not changed. 

On the other hand, the dynamical variables of the global particle are represented by 
operator fields of the form
$$
   Q : \chi \rightarrow Q_{\chi}  \quad .
$$
The expectation value of the dynamical variable $Q$ is given in terms of the statistical 
operator (\ref{eq.6.3}) by
\begin{equation}
 \label{eq.6.5}
    < Q >_K = \int_K Tr[Q_{\chi} \rho_{\chi} ] d\chi   \quad .
\end{equation}
If we use another measure on the dual, the operator field $Q$ will not be modified. This is due 
to the fact that $Q_{\chi}$ depends on the $\chi$ (the mass and spin, say) of the global 
particle, whereas $W$ 
has the additivity property characteristic of a measure: if $K_1$ and $K_2$ are two disjoint 
Borel parts of $\Gc$, one has
$$
   W(K_1 \cup K_2) = W(K_1) \oplus W(K_2)  \quad .
$$
$Q_{\chi}$ is analogous to physical quantities like the refraction index or the absorption 
coefficient, 
which are not densities of additive functions of the frequency interval, and which are 
therefore independent of the measure used on the frequency axis. 

The representation of a state by an operator-valued measure on the dual of the invariance group 
is not a familiar one; it is, however, unavoidable for a system like the global particle, which 
has the same dynamical variables as an elementary system(19), but is described by a reducible 
representation of the invariance group. 

Indeed, the definition of the global particle implies that its algebra of observables is 
isomorphic to that of an elementary system. Now the center of the universal enveloping algebra 
of the Lie algebra of the Poincar\'e  group is generated by the elements $P_{\mu}P^{\mu}$ and 
$W_{\mu}W^{\mu}$. For an elementary system, the representation of the group is irreducible and 
the operators which represent $P_{\mu}P^{\mu}$ and $W_{\mu}W^{\mu}$ are multiples of the 
identity operator. For a global particle the representation is no longer irreducible, it is 
only multiplicity free; but the operators which 
represent the elements of the center of the universal enveloping algebra still commute with all 
the dynamical variables (in other terms, they still belong to the center of the algebra of 
observables). Hence any state of a global particle is an incoherent superposition of states 
with fixed values of $\chi$. This is indeed precisely what is expressed by equation 
(\ref{eq.6.3}). 
Besides, if we take for $K$ the whole dual we get an unconditional statistical operator:
\begin{equation}
 \label{eq.6.6}
       W = \int_{\Gc}^{\oplus} \rho_{\chi} d\chi   \quad .
\end{equation}

In the just mentioned formalism of the algebra of observables, all observables are represented 
by operators on the same Hilbert space, which could be here only the unconditional space
\begin{equation}
 \label{eq.6.7}
    \mathcal{H} = \int_{\Gc}^{\oplus} \mathcal{H}_{\chi} d\chi   \quad .
\end{equation}
Such a representation is useful in certain cases, but here it would oblige us to assign to 
dynamical variables spectral densities, which are devoid of direct physical meaning. 

Let us point out at last that the formalism just introduced fits in well with the methods of 
experimental analysis of global particles, which have been used for instance in the study of 
strong processes. In such a case the measurements are conditioned by the choice of an interval 
of effective mass, and often also by the choice of a value for the total angular momentum of 
the system. This was especially conspicuous, for instance, in an experiment by Baton and 
Laurens(20): they studied dipions $\pi^- \pi_0$ in the mass region of the $\rho$ meson; their 
statistics was 
high enough to allow them conditioning their measurements by different mass intervals, whose 
width was most often smaller than that of the $\rho$ meson. But for any value of the mass width 
chosen, the measured dynamical variables are always functions of the mass of the global 
particle, whereas the elements of density matrix which can be measured depend additively of the 
mass interval. 

For those reasons the dynamical variables must be represented by operator fields, and the state 
must be represented by an operator-valued measure on the dual which allows conditional 
statistical operators such as (\ref{eq.6.3}) to be defined.

\subsection{Conservation of the Global Particle}

The value of the characteristic function of a particle (simple particle or global particle) at 
the element $g^{-1}$ of the invariance group is the expectation value of the operator of the 
representation of the invariance group. For a simple particle this operator is $\Ucg$; for 
a 
global particle it is given by the tensor product of the $n$ irreducible representations, i.e. 
by 
Eq. (\ref{eq.4.3}) (where one must put every $g_i$ equal to $g$), or equivalently by an 
operator field $\chi \rightarrow \Ucg$. Now the operators of the representation of the 
invariance group are conserved 
quantities, as are those which represent the elements of the Lie algebra (the former are the 
exponentials of the latter). Hence the expectation values of those operators are also conserved 
quantities. This implies that in any tranformation process, the global characteristic function 
of the system is conserved. We shall use in article II a particular case of this: if $n$ real 
particles transform into a virtual particle, the characteristic function of the virtual 
particle is equal to the global characteristic function of the initial $n$ particle system. 

For a process described by an $S$ matrix, one can give a formal proof of the conservation of 
the global characteristic function. Let $W_{in}$ and $W_{out}$ be respectively the ingoing and 
outgoing 
(normalized) statistical operators. Let $g-->U_g$ be the tensor product of the $n$ irreducible 
representations corresponding to the $n$ initial particles. The $S$ matrix commutes with the 
$U_g$ . The initial global characteristic function reads
$$
    \phi(in)_g = Tr(W_{in} U_{g^{-1}})
$$

The final characteristic function reads
\begin{equation}
 \begin{array}{llll}
 \nonumber
 \phi(out)_g &=& Tr(W_{out}U_{g^{-1}})& \\
             &=& Tr(SW_{in} S^{-1} U_{g^{-1}}) & \\
             &=& Tr(W_{in} S^{-1} U_{g^{-1}} S) & \\
             &=& Tr(W_{in} U_{g^{-1}})         &= \phi(in)_g  \quad . \\
 \end{array}
\end{equation}

\subsection{Some Details about the Meaning of Global Characteristic Functions}

In Ref. 9 it has been checked by elementary calculations (with use of the explicit forms of the 
matrix elements of the representations of $SU(2)$ and $P$, respectively) that the restrictions 
of 
the characteristic functions to the diagonal subgroups describe indeed the global state. These 
computations will not repeated here, but we shall now recall their main results. They show how 
the restriction of the characteristic function to the diagonal subgroup is reflected on the 
statistical operator, on the other side of the Fourier transformation. 

{\bf Example 1: $SU(2)$}

Let us consider first the case of two spins $j_1$ and $j_2$. The complete state of the system 
is 
described by a statistical operator $W$ which can be written either in the basis of individual 
states $|m_1m_2>$, or in the basis of global states $|JM>$. To go from one basis to the other 
one 
one uses the Clebsch-Gordan coefficients $<m_1m_2|JM>$. In the basis $|JM>$, the operator $W$ 
is not diagonal in general. To pass from $W$ to $W_{glob}$, we must eliminate the elements 
which are non-diagonal with respect to $J$ :
\begin{equation}
 \label{eq.6.8}
    W_{glob} = \oplus W_J
\end{equation}
with
\begin{equation}
 \label{eq.6.9}
   <M | W_J | M'> = <JM | W | JM'>
\end{equation}
Otherwise stated, the operator $W_glob$ is obtained from $W$ by diagonal truncation with 
respect to 
$J$ : if we call $\Pi_J$ the projector onto the subspace defined by the value $J$ for the total 
angular momentum, one has
\begin{equation}
 \label{eq.6.10}
    W_J = \Pi_J W \Pi_J
\end{equation}

If one computes the characteristic function $f$ corresponding to the operator (\ref{eq.6.8}):
\begin{equation}
 \label{eq.6.11}
   f_g = \sum_J Tr [W_J U_J(g^{-1}) ] \quad ,
\end{equation}
one can see that it is equal to the diagonal restriction of the complete characteristic function. 

Let us now consider the case of more than two spins. We have to do with a degenerate case: the 
common eigenspaces of $J^2$ and $J_z$ are now more than one-dimensional. We shall denote the 
vectors 
of the collective basis by $|JMa>$, where $a$ stands for one or several degeneracy parameters. 
The vectors of the individual basis will be denoted by $|m_1...m_n>$, or $|(m_i)>$ for short. 
The equations analogous to (\ref{eq.6.8})-(\ref{eq.6.9}) read:
\begin{equation}
 \label{eq.6.12}
    W_{glob} = \oplus W_J
\end{equation}
with
\begin{equation}
 \label{eq.6.13}
   <M | W_J | M'> = \sum_a <JMa | W | JM'a>  \quad .
\end{equation}
Here going from $W$ to $W_J$ is more complicate: equation (\ref{eq.6.10}) is replaced by
\begin{equation}
 \label{eq.6.14}
    W_J = Tr_a [\Pi_J W \Pi_J]  \quad .
\end{equation}
We have taken the partial trace with respect to the degeneracy parameters $a$.

Here again, one can check explicitly that the characteristic function corresponding to the 
operator (\ref{eq.6.12}) is equal to the diagonal restriction of the complete characteristic 
function.

{\bf Example 2: Poincar\'e group}

The main difference with the $SU(2)$ case is due to the fact that the dual is no longer 
discrete. 
In the definition of the global statistical operator $W_{glob}$, the direct sum of Eq. 
(\ref{eq.6.8}) is replaced by a direct integral(8)(13):
\begin{equation}
 \label{eq.6.15}
       W_{glob} = \int_{\Gc}^{\oplus} \rho_{\chi} d\chi   \quad .
\end{equation}
To make explicit the definition of $\rho_{\chi}$, we shall use continuous pseudobases in the 
spaces $\Hc_{\chi}$. We 
shall use the results and the notations of Joos(21). Let us take the case $n = 2$. A one 
particle 
state will be denoted by $|k>$, where $k$ stands for both the energy-momentum and the spin or 
helicity index. Similarly, a two particle state will be denoted by $|k_1k_2>$. As we have 
almost 
always a degenerate case (the only exception is the case where at least one of the spins is 
zero), a global state will always be denoted by $|\chi k \eta>$, where $\eta$ stands for one or 
several 
degeneracy parameters. The Clebsch-Gordan coefficients (whose explicit expressions are given by 
Joos) will be denoted by  $<k_1k_2 | \chi k \eta>$. The expression of an element of the global 
pseudobasis is
$$
| \chi k \eta> = \sum_{sp}\int d^3 k_1 [2w_1]^{-1} d^3 k_2 [2w_2]^{-1} | k_1k_2 ><k_1k_2 | \chi k \eta >
$$
 where one has put
$$
  w_i = (\mathbf{k}_i^2+m_i^2)^{1/2}  \quad  ;
$$
here $\mathbf{k}$ stands for the space part of $k$; 
$\sum_sp$ denotes summation over the spin indices contained in $| k_1k_2>$.
 
Let us call, as in Eq. (\ref{eq.6.5}), $\rho$ the operator field "density of global statistical 
operator". 
The (improper) matrix elements of $\rho_{\chi}$ can be expressed in terms of those of the 
complete statistical operator $W$:
\begin{equation}
 \label{eq.6.16}
       <k |\rho_{\chi} | k'> = \sum_{\eta} < \chi k \eta | W |\chi k' \eta >  \quad .
\end{equation}

As shown by equation (\ref{eq.6.2}), the characteristic function corresponding to the 
operator-valued 
measure whose density is $\rho_{\chi}$ is the inverse Fourier transform of the field $\rho$. It 
can be checked 
explicitly that the latter is equal to the diagonal restriction of the complete characteristic 
function. 

The cases $n>2$ differ from the case $n=2$ only by some complications; $\eta$ becomes a 
continuous 
parameter. The Clebsch-Gordan coefficients for any $n$ have been computed by Klink and 
Smith(22)(23). 

Let us point out that all the results stated about the global state and the global 
characteristic function follow from the sole assumption of relativistic invariance. The 
properties of these objects are elementary kinematical properties. Their meaning did not appear 
clearly in Ref. 8 and Ref. 9, where they were mixed with dynamical considerations. 

ACKNOWLEDGMENTS

F.L. wishes to thank Gilbert Arsac, Pierre Bonnet and Pierre Eymard for useful mathematical informations and discussions. 

APPENDIX: FUNCTIONAL SPACES  

We have introduced two types of characteristic functions. The functions which describe one or 
several particles are the inverse Fourier-Stieltjes transforms of Dirac measures on the dual of 
a group (the invariance group, or a direct power of it); the functions which describe global 
particles are the inverse Fourier transforms of operator fields on the dual of the invariance 
group. Let us show that these two types of functions belong respectively to two types of 
functional spaces. These spaces, which play an important role in harmonic analysis, have been 
introduced by Eymard(24) in 1964. (For a recent review, see Eymard(25)). 

The Fourier-Stieltjes algebra $B(G)$ of a locally compact group $G$ is the vector space of the 
functions over $G$ defined by: $g \rightarrow <x|U_g|\eta>$ , where $U$ is a representation 
(not necessarily 
irreducible) of the group. This space is also a Banach algebra, for a product and a norm which 
need not be specified here. 

The Fourier algebra $A(G)$ of a locally compact group $G$ is the closure in $B(G)$ of the 
intersection of $B(G)$ with the space $L(G)$ of the continuous functions with compact support 
on $G$. All the functions of $A(G)$ tend to zero at infinity (Ref. 22, proposition (3.7)). 

These definitions may look somewhat abstract; they will become more familiar by specifiying the 
relations of $A(G)$ and $B(G)$ with the operator fields and the operator-valued measures on the 
dual.

Let $\mathcal{L}_1(G)$ be the vector space of the trace class operator fields on $\widehat{G}$, 
 i.e. 
of the operator fields  $\chi \rightarrow A_{\chi}$  such that the function             
$\chi \rightarrow Tr(|A_{\chi}|)$ exists and 
is integrable. The relation of almost everywhere equality (for the Plancherel measure) is an 
equivalence relation, and the space of classes modulo this relation is denoted by $L_1(G)$. We 
shall say for simplicity's sake that $L_1(G)$ is the space of integrable fields on $G$. 

It follows immediately from the definition that an integrable field has an inverse Fourier 
transform:
$$
\mathcal{T}^{-1}(A)_g = \int_G Tr[A_{\chi} \Uc{g-1}] d\chi   \quad .
$$

Let $\mathcal{M} = (\mu,F)$ be a trace class operator-valued measure on $\Gc$ (see Section 
IIb), and let 
$M(G)$ be the space of all trace class operator-valued measures. We know that $\mathcal{M}$ 
has an inverse Fourier-Stieltjes transform. 

We can now characterize $A(G)$ and $B(G)$ by using Fourier transforms. The Fourier algebra 
$A(G)$ 
is the image by the inverse Fourier transformation of the space $L_1(G)$. The Fourier-Stieltjes 
algebra $B(G)$ is the image by the inverse Fourier-Stieltjes transformation of the space 
$M(G)$. 

It follows then from Section 4 that the characteristic function of an $n$ particle state ($n \geq  1$) 
belongs to the Fourier-Stieltjes algebra of the $n^{th}$ direct power of the invariance group. 

According to Eq. (\ref{eq.6.2}) the global characteristic functions are the inverse Fourier 
transforms of elements of $L_1(G)$. Hence they belong to the Fourier algebra of the 
invariance group. 

The difference between the two cases is due to the fact that the characteristic function of an 
$n$ particle state corresponds to fixed values of the masses; on the contrary, a global 
particle has a continuous mass spectrum. 

We can see thus that the Fourier and Fourier-Stieltjes algebras 
might have been tailor-made for quantum mechanics.

REFERENCES
\begin{list}%
{\arabic{num}}{\usecounter{num}}

\item J.-Y.Grandpeix, F.Lur\c{c}at,{\it Particle Description of Zero Energy Vacuum, II. Basic 
Vacuum Systems}, following article.
\item S.Weinberg, {\it The Quantum Theory of Fields}, vol.I (Cambridge University Press, 
Cambridge, 1995). \item J.Frenkel, {\it Doklady Akademii Nauk SSSR} 64, 507 (1949).
\item W.E.Thirring, {\it Principles of Quantum Electrodynamics} (Academic Press, New York, 
1958). \item L.Rosenfeld, {\it Nucl. Phys.} 10, 508 (1959).
\item S.S.Schweber, {\it Rev. Mod. Phys.} 58, 449-509 (1986).
\item S.S.Schweber, {\it QED and the Men Who Made It: Dyson, Feynman, Schwinger, and Tomonoga} 
(Princeton University Press, Princeton, 1994).
\item F.Lur\c{c}at, {\it Phys. Rev.} 173, 1461 (1968).
\item F.Lur\c{c}at, {\it Ann. Phys.} (N.Y.) 106, 342 (1977).
\item N.Wiener, {\it The Fourier Integral and Certain of its Applications} (Cambridge 
University Press, Cambridge, 1933; republished by Dover, New York).
\item E.Hewitt, K.A.Ross, {\it Abstract Harmonic Analysis}, vol. I (Springer, Berlin, 1963); 
vol. II (Springer, Berlin, 1970).
\item A.A.Kirillov, {\it Elements of the Theory of Representations}, translated from Russian by 
E.Hewitt (Springer, Berlin, 1976)
\item J.Dixmier, {\it C*-Algebras} (North-Holland, Amsterdam, 1977).
\item P.Bonnet, {\it J.Funct.Anal.} 55, 220-246 (1984).
\item Nghi\^em Xu\^an Hai, {\it Commun. math. Phys.} 12, 331-350 (1969).    
\item Nghi\^em Xu\^an Hai, {\it Commun. math. Phys.} 22, 301-320 (1971).
\item A.M.Yaglom, {\it Second-order homogeneous Random Fields, Proc. IVth Berkeley Symposium Math. 
Statistics and Probability}, vol.2, 593-622 (University of California Press, Berkeley,1961).
\item U.Fano, {\it Rev. Mod. Phys.} 29, 74-93 (1957).
\item T.D.Newton, E.P.Wigner, {\it Rev. Mod. Phys.} 21, 400-406 (1949).
\item J.P.Baton, G.Laurens, {\it Phys. Rev.} 176, 1574-1586 (1968).
\item H.Joos, {\it Fortschritte der Physik} 10, 65-146 (1962).
\item W.H.Klink, G.J.Smith, {\it Commun. math. Phys.} 10, 231-244 (1968).
\item W.H.Klink, {\it Induced Representation Theory of the Poincar\'e  Group}, in {\it "Mathematical 
Methods 
in Theoretical Physics"}  (Boulder, Colorado, 1968; K.T.Mahantappa, W.E.Brittin, eds.), 
{\it Lectures in Theoretical Physics}, vol. 11D (Gordon \& Breach, New York, 1969).
\item P.Eymard, {\it Bull. Soc. Math. France} 92, 181-236 (1964).
\item P.Eymard, {\it A Survey of Fourier Algebras}, in {\it "Applications of Hypergroups and Related 
Measure Algebras"}  (Seattle, Washington, 1993; W.C.Connett, M.-O.Gebuhrer, A.L.Schwartz, 
eds.), {\it Contemporary Mathematics}, 183, 111-128 (American Mathematical Society, Providence, 
Rhode Island, 1995).

\end{list}

\end{document}